\documentclass[sigconf]{acmart}
\usepackage{placeins}
\usepackage{float}

\AtBeginDocument{%
  \providecommand\BibTeX{{%
    \normalfont B\kern-0.5em{\scshape i\kern-0.25em b}\kern-0.8em\TeX}}}

\setcopyright{rightsretained}
\copyrightyear{2019} 
\acmYear{2019} 
\acmConference[Koli Calling '19]{19th Koli Calling International Conference on Computing Education Research}{November 21--24, 2019}{Koli, Finland}
\acmBooktitle{19th Koli Calling International Conference on Computing Education Research (Koli Calling '19), November 21--24, 2019, Koli, Finland}
\acmDOI{10.1145/3364510.3366159}
\acmISBN{978-1-4503-7715-7/19/11}

\begin{document}

\title{The Effect of Civic Knowledge and Attitudes on CS Student Work Preferences}

\author{Antti Knutas}
\affiliation{%
  \institution{LUT University}
  \city{Lappeenranta}
  \country{Finland}
}
\email{antti.knutas@lut.fi}
\orcid{0000-0002-6953-0021}

\author{Andrew Petersen}
\affiliation{%
  \institution{University of Toronto}
  \city{Toronto}
  \country{Canada}
}
\email{andrew.petersen@utoronto.ca}
\orcid{0000-0003-1337-7985}

\renewcommand{\shortauthors}{-}

\begin{abstract}
We present an investigation in the connection between computing students' civic knowledge, attitude, or self-efficacy and their willingness to work on civic technologies. Early results indicate that these factors are related to a willingness to accept government work in technology but not non-profit work focused on civic technologies.
\end{abstract}



\maketitle

\section{Introduction}
Advanced e-government initiatives are ongoing in many countries in the world, where government services are being offered online in decentralized platforms~\cite{yli2018suomi}. There are many ways to participate in the development of civic technology~\cite{palacin2018understanding}, or technology for social change~\cite{patel2013emergence}. While some studies have been performed in the career choice awareness of computer science graduates \cite{calitz2011ict}, there are few studies about their interest in participating in the development of government-driven civic technology (govtech~\cite{gilman2017civic}). By comparison, volunteer work such as civic hackathons~\cite{johnson2014civic} or CS education for social good~\cite{goldweber2013framework} have been better explored in literature.

These initiatives require skilled technology professionals to enable the next generation of technologies. However, there is a competition for the most skilled professionals in the computing and software engineering job market. Work placements have been shown to be beneficial to careers after graduation~\cite{brooks2016undergraduate} and learning about the industry affects the students' career choices \cite{scott2009more}.

What remains mostly unexplored is what kind of attitudes and knowledge computer science students have about civics and if this knowledge affects their career choices. To address this gap, we performed a survey-based research study on students' civic knowledge, attitude and self-efficacy, and their willingness to work on civic technologies. In summary, our research question is: \textit{Do civic knowledge, attitude, or self-efficacy affect the work preferences of students?}
 

In the following sections, we first first describe the research setup, data collection, and methods. Then, we review highlights of our early findings.

\section{Research Setup}
Data was collected from the second week of a 12 week CS1 course at the University of Toronto, a research-intensive North American university. Students in the course are primarily first year university students interested in studying computer science, mathematics, or statistics. 

\begin{table*}[h!tp]
    \centering
    \begin{tabular}{p{3in}lllllll}
         & Strongly  & & Slightly &  Neither Agree  & Slightly & & Strongly\\
         Question & Disagree & Disagree & Disagree & nor Disagree & Agree & Agree & Agree\\\hline
         1. Participating in activities to benefit people in my local community is important.& 2 & 7 & 3 & 23 & 28 & 68 & 17\\
         2. I am interested in developing civic technologies (e.g., technologies that enhance the relationship between citizens and their government).& 3 & 10 & 9 & 35 & 36 & 39 & 16\\ 
         3. I would participate in a hackathon or other event to develop civic technologies.& 4 & 9 & 2 & 23 & 37 & 50 & 23\\
         4. I would volunteer my time and technical skills to contribute to the development of civic technologies.& 3 & 8 & 8 & 28 & 46 & 36 & 19\\
         5. I would work for a non-profit that develops civic technologies.& 3 & 16 & 12 & 41 & 45 & 20 & 11\\
         6. I would work for the government.& 10 & 14 & 14 & 37 & 26 & 30 & 17\\
    \end{tabular}
    \caption{Civic Technologies Survey Questions and Responses} 
    \label{table:mc}
    \vspace*{-1em}
\end{table*}

Two surveys were made available to all of the students in the course. Both were optional, with no material reward for participating. The first contained six Likert-style questions about a student's willingness to engage in the development of civic technologies. Table~\ref{table:mc} lists the questions in the survey; students responded on a 7-point scale from ``strongly disagree'' to ``strongly agree''. 148 students provided responses to all six questions in this survey. Respondents were primarily male (74\%), which reflects the make-up of the course. The second survey was the ICCS civic knowledge, attitude, and self-efficacy survey instrument \cite{schulz2010iccs}. 52 students participated in the ICCS survey; the lower participation rate is likely due to the length of the instrument.

\subsection{Analysis Methods}
In order to analyze the effect of civic knowledge and engagement on willingness to participate in the development of civic technologies and to work in the government sphere, we applied ordinal logistic regression~\cite{o2006logistic} due to our use of Likert-scale items. We evaluated multiple possible models and used the Akaike Information Criterion (AIC) and Schwarz Criterion (SC) to identify models of interest.

There is no default significance test for coefficients; we estimate a p-value by comparing the coefficient's t value to a normal distribution.

\section{Findings}

Table~\ref{table:mc} describes the responses to the six questions about civic technologies. It is heartening that a significant majority of the students were willing to agree that they were interested in civic technologies. We compared the students who agreed with the first question to those who agreed to the fourth, and found that almost all who were interested in civic technologies also agreed that they would be willing to volunteer their time. A larger number of students were willing to attend a hackathon with an emphasis on civic technologies, which suggests that limited term events may be effective at exposing students to the domain.

A much smaller number of respondents were willing to work on civic technologies, either at a non-profit or in the government. We do not have any data to explain why volunteering is more appealing than working, but we speculate that students may associate work at a non-profit or government with less interesting or cutting-edge technologies or a lower rate of pay. This question will be the focus of future work with this population.  

We considered whether gender might impact responses. However, using visual inspection of aggregate responses for males and females, we saw no change in the trends reported above.

Finally, we will consider how the components of the ICCS survey are related to engagement with civic technologies. The ICCS survey has a number of components including, for example, perception of conventional citizenship activities, attitudes towards rights for minority elements of the population, and trust in civic institutions \cite{schulz2010iccs}. Due to space restrictions, we will focus on models of willingness to work at a non-profit developing civic technologies or to work for the government. 

Willingness to work at a non-profit is interesting because no models were found that included significant coefficients. The better (low AIC and SC) models had no coefficients that were estimated to be significant at even the 0.05 level. In particular, not even the questions related to interest in social issues or equality were significant to the model. This result could be caused by the relatively small amount of data collected, but it could also indicate that other factors, such as financial goals, might dominate the decision to consider employment at a non-profit.

In contrast, the best model for willingness to work for the government had a number of significant coefficients, including questions related to trust in civic institutions and attitudes toward country. Again, questions about political or social issues are noticeably absent, as are questions about conventional citizenship activities, like voting and affiliating with a party.

\section{Conclusion}
In this paper, we presented our initial findings on the impact of civic knowledge, attitude, or self-efficacy on the work  preferences of students.

The initial findings of our study indicate that civic knowledge or attitudes -- including attitudes related to social issues and equality -- do not have a significant effect on the choice to work at a non-profit. When it comes to working for the government, trust in civic institutions and attitudes towards country were significant coefficients.

Our findings suggest that educational and awareness efforts alone might not be enough to make civic work an interesting career choice to computing professionals after graduation. However, at the same time the students were interested in creating civic technology as volunteer work.

The findings are not yet conclusive and more data needs to be collected before the results can be used to support decision-making for practitioners. However, the initial results presented in this paper can inform future studies in the field.

\begin{acks}
We thank our colleague Victoria Palacin from LUT University on sharing her expertise on the field of civic technology.
\end{acks}

\FloatBarrier

\bibliographystyle{ACM-Reference-Format}
\bibliography{references}

\end{document}